\documentclass[traditabstract]{aa}
\usepackage[utf8]{inputenc}
\usepackage{amsmath}
\usepackage{amssymb}
\usepackage{microtype}
\usepackage{natbib}
\usepackage{graphicx}
\usepackage{txfonts}
\usepackage{color}
\usepackage{multirow}
\usepackage{xspace}
\usepackage{ulem}
\bibpunct{(}{)}{;}{a}{}{,} % to follow the A&A style
\newcommand{\src}{\object{SGR~0755$-$2933}\xspace}
\newcommand{\srcnew}{\object{2SXPS~J075542.5$-$293353}\xspace}

\begin{document}
\title{SGR~0755$-$2933: a new High Mass X-ray binary with the wrong name}
\author{V.\,Doroshenko\inst{1,2} \and A.\, Santangelo\inst{1}  \and S. S.~Tsygankov\inst{3,2} \and L.\, Ji\inst{1}}
\institute{Institut für Astronomie und Astrophysik, Sand 1, 72076 Tübingen, Germany
\and
Space Research Institute of the Russian Academy of Sciences, Profsoyuznaya Str. 84/32, Moscow 117997, Russia
\and 
Department of Physics and Astronomy,  FI-20014 University of Turku, Finland}

\authorrunning{V. Doroshenko et al.}
\abstract{The soft gamma-ray repeater candidate \src was discovered in 2016 by \textit{Swift}/BAT, which detected a short ($\sim30$\,ms) powerful burst typical of magnetars. 
To understand the nature of the source, we present here an analysis of follow-up observations of the tentative soft X-ray counterpart of the source obtained with \textit{Swift}/XRT, \textit{NuSTAR} and \textit{Chandra}. From our analysis we conclude that, based on the observed counterpart position and properties, it is actually not a soft gamma ray repeater but rather a new high mass X-ray binary. We suggest to refer to it as 2SXPS~J075542.5$-$293353. We conclude, therefore, that the available data do not allow us to confirm existence and identify the true soft X-ray counterpart to the burst event. Presence of a soft counterpart is, however, essential to unambiguously associate the burst with a magnetar flare, and thus we conclude that magnetar origin of the burst and precise burst location remain uncertain and require further investigation.}
\maketitle

\section{Introduction}
A new Soft Gamma-ray Repeater (SGR) \src candidate was
discovered on Mar 16 2016 following a short ($\sim30$\,ms), soft
burst triggered by Burst Alert Telescope on board the Neil Gehrels Swift Observatory (\textit{Swift}/BAT)
around RA=118.884, Dec=$-$29.552 (3$\arcmin$ error radius, 90\%  containment) and subsequently refined by \cite{gcn19207} to RA=118.8625, Dec=$-$29.5723 (with 3.4$\arcmin$ error radius, 90\%  containment). 
Considering that the SGRs are scarce and believed to be powered by neutron star crust fractures associated with magnetic field re-arrangements in magnetars \citep{1995MNRAS.275..255T}, this immediately triggered a follow-up campaign with multiple facilities.
The tentative soft X-ray counterpart was identified at RA=118.9270 Dec=-29.5637 (2$\arcsec$ error radius, 90\% containment), minutes after the burst following an automated re-pointing of \textit{Swift}/XRT \citep{2016ATel.8831....1B}.
Further monitoring
of the transient with \textit{Swift}/XRT showed a slow and steady fading of the source in the 0.5-10\,keV band from
\hbox{$\sim8\times10^{-11}$\,erg\,cm$^{-2}$\,s$^{-1}$} (on MJD 57463.9) to \hbox{$\sim2\times10^{-11}$\,erg\,cm$^{-2}$\,s$^{-1}$} (on MJD 57464.5). However, no further bursts nor X-ray pulsations in the frequency range typical for magnetars were detected \citep{2016ATel.8868....1B}.
Observations in the radio band from 327 and 1390~MHz \citep{2016ATel.8943....1B} to 2~GHz and 6~GHz \citep{2017AAS...22943104H} also did not result in a significant detection of pulsations. \cite{2017AAS...22943104H} mentioned a hint of a $\sim300$\,s pulsations in the X-ray band, however, no details were reported. Furthermore, the source has been observed by \textit{Chandra} and \textit{NuSTAR} on several occasions, but to the best of our knowledge, no results from these observations have been reported in the literature. 

In this paper, we focus on \textit{Chandra}, \textit{NuSTAR} and \textit{Swift}/XRT observations of the suggested soft X-ray counterpart to this event. We conclude that the position and the X-ray properties of the suggested counterpart indicate that this is not a soft gamma repeater but rather a new Be X-ray binary (BeXRB), \srcnew, which falls within the error circle for the burst most likely by chance.
On the other hand, other X-ray sources detected within its reported error circle are too faint for the detailed analysis required to confirm the presence of a putative magnetar. 
 The localization and classification of the burst as a soft gamma repeater remain, therefore, uncertain as suggested presence of the soft counterpart was the main argument in favour of its magnetar origin.

Faintness of other potential counterparts to the burst, precludes detailed analysis of their properties, so here we focus instead on the discussion of the observed properties of the newly identified  high-mass X-ray binary (HMXB), \srcnew. We confirm the presence of $\sim308$\,s pulsations and, based on the observed long-term variability, identify the tentative orbital period of $\sim260$\,d. Considering the observed spin and orbital period of the source and the classification of the optical companion as an emission line star, we finally conclude that the system is likely a new low-luminosity  BeXRB with $L_x\sim10^{34}$\,erg\,s$^{-1}$ at $\sim3.5$\,kpc. This conclusion is in line with the observed two-component broadband X-ray spectrum of the source similar to that observed in other low luminous BeXRBs.

\section{Observations and data analysis}
\begin{table}
	\begin{center}
	\begin{tabular}{ccccc}
	  Instr. & Obsid & Date (MJD) & Exp.(ks) & Group \\
		\hline
		 N & 80102103002 & 57473.20 & 54.7 &A \\
		C & 18014 & 57474.08 & 10 & A \\
		 N & 80102104002 & 57497.37 & 76.2 &B \\
		C & 18015 &  57500.48 & 15 & B\\
		N & 80102103004 & 57589.09 & 28.7 & C\\
		N & 80102104003 & 57589.90 & 107.8& C \\
		C & 18016 & 57592.46 & 25 & C \\
		C & 18017 & 57752.06& 40 &D \\
		C & 22454 & 58759.21 & 30 &E\\
		 N & 90601322001 & 59049.49 & 49.8 &A\\
		\hline
		
	\end{tabular}
	\end{center}
	\caption{List of dedicated \textit{NuSTAR} (N) and \textit{Chandra} (C) observations of the source considered in this work (grouped by observed flux).}
	\label{tab:obs}
\end{table}
The list of \textit{NuSTAR} and \textit{Chandra} observations used in this work is
presented in Table~\ref{tab:obs}. The data reduction was performed according to each instrument's documentation with the help of \texttt{HEASOFT v6.28} (CALDB 20200912)
for \textit{NuSTAR} and \texttt{CIAO 4.12} (CALDB 4.9.2.1) for \textit{Chandra}. Data from the two instruments were used primarily for broadband spectral and timing analyses (\textit{NuSTAR}), and source localization (\textit{Chandra}). We also analyzed 45 available \textit{Swift}/XRT observations with a total exposure of $\sim140$\,ks to investigate the long-term variability of the source. Analysis of the \textit{Swift}/XRT data was performed using online tools\footnote{\url{http://www.swift.ac.uk/user_objects/}} \citep[][]{2009MNRAS.397.1177E} provided by the UK Swift Science Data Centre. In particular, source spectra for each observation were extracted using the standard extraction options to assess the long-term flux variations.

\subsection{Source localization and counterpart}
\begin{figure}
    \centering
    \includegraphics[width=\columnwidth]{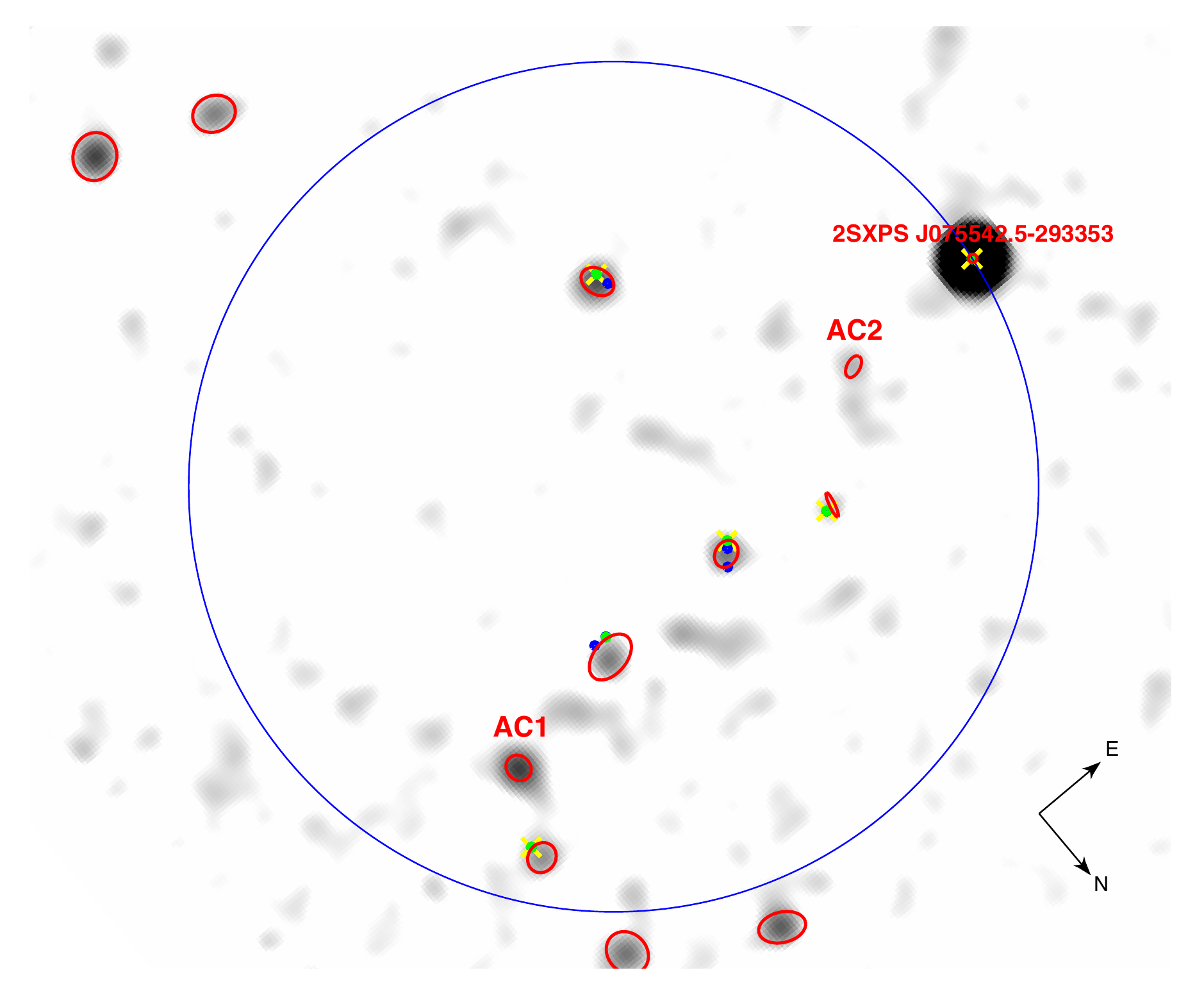}
    \caption{\textit{Chandra} broad (0.5-7\,keV) band flux image covering \textit{Swift}/BAT 90\% error circle for the burst as reported in \cite{gcn19207} (blue circle). Besides the originally suggested counterpart (labeled here \srcnew) several faint X-ray sources (red ellipses) are detected by \textit{Chandra} within BAT confidence region and just outside of it. At least two of those (labeled AC1/2) appear to have no obvious optical (Gaia DR2, blue circles) or near infrared (2MASS, green circles, AllWise, yellow crosses) counterparts and thus, we suggest, could be true soft X-ray counterparts to the burst detected by BAT.}
    \label{fig:map}
\end{figure}

The position of \src was already rather well determined by XRT \citep{2016ATel.8831....1B}. However, the most accurate localization can be obtained using the \textit{Chandra} imaging-mode data (obsid. 22454). To do that, we constructed a fluxed image of the field in broad 0.5-7\,keV energy range. This choice is justified by relatively low counting statistics and the fact that spectra of all known magnetars are relatively hard. Source-detection was then performed as prescribed in the \textit{Chandra} documentation (including the absolute astrometric correction with the 2MASS catalogue as a reference).
As a result, 37 point sources in total were detected. Six of them lie within the BAT error circle as shown in Fig.~\ref{fig:map}. The position of the brightest source among the six ($\sim2400$ counts) is consistent with the XRT localization of the magnetar candidate, i.e., RA=118.9271642 Dec=$-$29.5648607 (07 55 42.5194,  $-$29 33 53.4985) and $\sim2\arcsec$ uncertainty (at $3\sigma$ confidence level, including 1.5$\arcsec$ systematic). This is the only source for which a detailed analysis is feasible since only $\sim90$ counts are detected from the second brightest source and even less from the rest, so below we mostly focus on analysis of properties of this object.

First of all, it is important to note it has a bright infrared and optical counterpart with \textit{J}$\sim9.7^m$, \textit{G}$\sim9.94^m$ detected at an offset $\sim0.6\arcsec$ in the 2MASS and Gaia surveys. This object, CD-29~5159, is a known luminous early type (O6) emission-line star \citep{1971PW&SO...1a...1S,2003AJ....125.2531R}, which strongly suggests that initial classification of corresponding X-ray source as a magnetar and soft gamma ray repeater \src is likely incorrect. Therefore, to avoid confusion, we refer to this source as \srcnew through the rest of the manuscript.  
We emphasize, however, that the arguments discussed by \cite{2016ATel.8831....1B} in favor of magnetar origin for the burst itself are still valid, so one could still expect to detect a soft X-ray counterpart to this event. Indeed, there are several fainter objects detected by \textit{Chandra} within the \textit{Swift}/BAT error circle as shown in Fig.~\ref{fig:map} which are also listed in the Table~\ref{tab:det}. All sources with an exception of two (marked as AC1/AC2) in the table), however, appear to have potential counterparts in 2MASS/Gaia surveys and thus are likely field stars or active galactic nuclei (AGNs). 

We emphasize that absence of an obvious near infrared or optical counterpart alone can only be viewed as a \textit{tentative} indication of a possible isolated neutron star origin, and and only due to the fact that other X-ray sources appear to have an optical counterpart which is not expected for a magnetar. There is, therefore, no strong indication that this object is indeed a magnetar, and our suggestion can only be verified through detailed analysis of X-ray properties of the source, for instance through detection of X-ray pulsations. Unfortunately, with only 10-30 net counts, and a detection significance of $\sim2-7\sigma$ (estimated by \texttt{wavdetect} task) both objects are too faint for detailed analysis. We conclude, therefore, that deeper observations both in X-ray and optical bands are required to unambiguously identify true counterpart to the burst detected by BAT, which is beyond the scope of current investigation. 
Instead, we focus exclusively on the analysis of \srcnew and demonstrate that it is indeed a new low-luminous HMXB.
\begin{table}[t]
    \centering
    \begin{tabular}{llll}
RA (J2000) & DEC (J2000)  & Net (bgd) counts & signif. ($\sigma$)\\
\hline
07 55 42.528 &  $-$29 33 53.64 & 2442.3 (34.7) & 351.33\\
07 55 31.392 &  $-$29 35 40.56 & 34.6 (8.4) & 8.62\\
07 55 17.664 &  $-$29 33 06.12$^{AC1}$ & 34.0 (12.0) & 7.44\\
07 55 28.584 &   $-$29 33 20.88 & 28.0 (8.0) & 7.06\\
07 55 22.872 &  $-$29 33 18.72 & 26.3 (7.7) & 6.74\\
07 55 16.224 &  $-$29 32 25.8 & 19.4 (8.6) & 4.79\\
07 55 36.6 &  $-$29 33 50.4$^{AC2}$ & 9.8 (5.2) & 2.86\\
07 55 32.736 &  $-$29 33 06.12 & 5.5 (1.5) & 2.20\\ 

\end{tabular}
    \caption{List of \textit{Chandra} sources within \textit{Swift/BAT} error circle. Alternative candidate counterparts to \src are marked as AC1/AC2. Estimated source net and background fluxes and the detection significance are reported as estimated by \texttt{wavdetect} task.}
    \label{tab:det}
\end{table}

\subsection{Timing analysis}

\begin{figure}%[t!]
    \centering
    \includegraphics[width=\columnwidth]{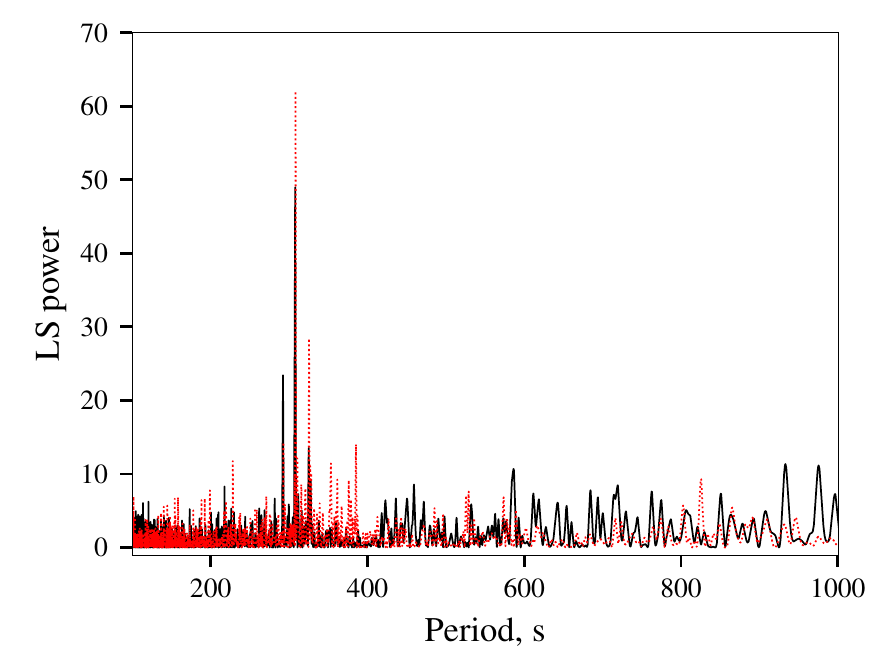}
    \includegraphics[width=\columnwidth]{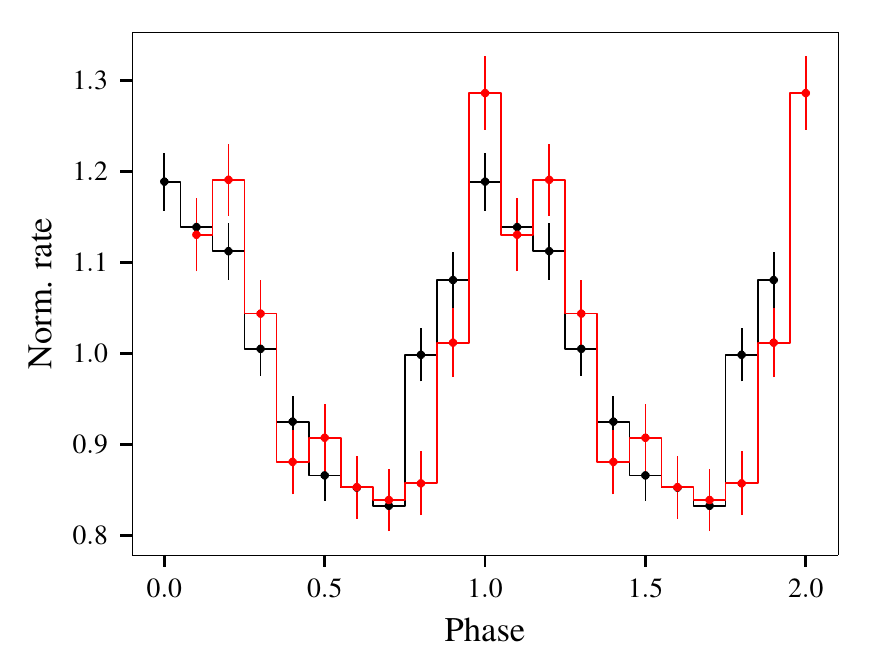}
    \caption{Lomb-Scargle periodogram and folded pulse profiles
    in 3-80\,keV energy range. Only results for the two \textit{NuSTAR} observations where the source is brightest 80102103002 (black) and 90601322001 (red) are shown for clarity.}
    \label{fig:peri_pp}
\end{figure}

As mentioned by \cite{2017AAS...22943104H}, X-ray data suggests that \srcnew pulsates  with period of $\sim308$\,s, which would be unprecedented for magnetars but typical of accretion-powered pulsars. To confirm the pulsing nature of the source, we extracted lightcurves in the 3-80\,keV energy band and searched for pulsations in the range 1-1000\,s in all \textit{NuSTAR} observations. To extract the lightcuves and spectr we followed the recommendations available on the online threads for each instrument.
In particular, the source and background spectra and light-curves were extracted from circular regions with radius of 55$\arcsec$ and 100$\arcsec$ centered on the source and on a source-free region respectively. The extraction radius was optimized to improve the signal to noise ratio above 20\,keV as described in \cite{2018A&A...610A..88V}. For each observation, data from the two \textit{NuSTAR} telescope units were analysed independently. 

A strong signal with period around 300\,s is indeed present in all cases in the Lomb-Scargle periodogram of the source (see upper panel of Fig.~\ref{fig:peri_pp}). The spin period is measured to be 307.80(4)\,s and 308.26(2)\,s for the first and last observation respectively (obsid. 80102103002 and 90601322001, the period value and uncertainties are estimated using the method described by \hbox{\citealt{2013AstL...39..375B})}. The slight increase of the spin period may indicate a spin-down trend, or may as well be due to the orbital motion since the source's orbit is not known. The observed pulse profiles in both cases exhibit a broad single peak and pulsed fraction\footnote{Calculated as $(F_\mathrm{max}-F_\mathrm{min})/(F_\mathrm{max}+F
_\mathrm{min})$, where $F_\mathrm{max}$ and $F_\mathrm{min}$ are maximum and
minimum fluxes in the pulse profile, respectively.} of 19(2) and 22(2)\% respectively (after subtraction of the background). The statistical quality of the available data precludes a more detailed analysis as illustrated in Fig.~\ref{fig:peri_pp}. Nevertheless, we note that both pulse profile shape and pulsed fraction are similar to those observed in other pulsars accreting at low luminosities \citep{2019MNRAS.487L..30T}.

\begin{figure}%[t!]
    \centering
    \includegraphics[width=\columnwidth]{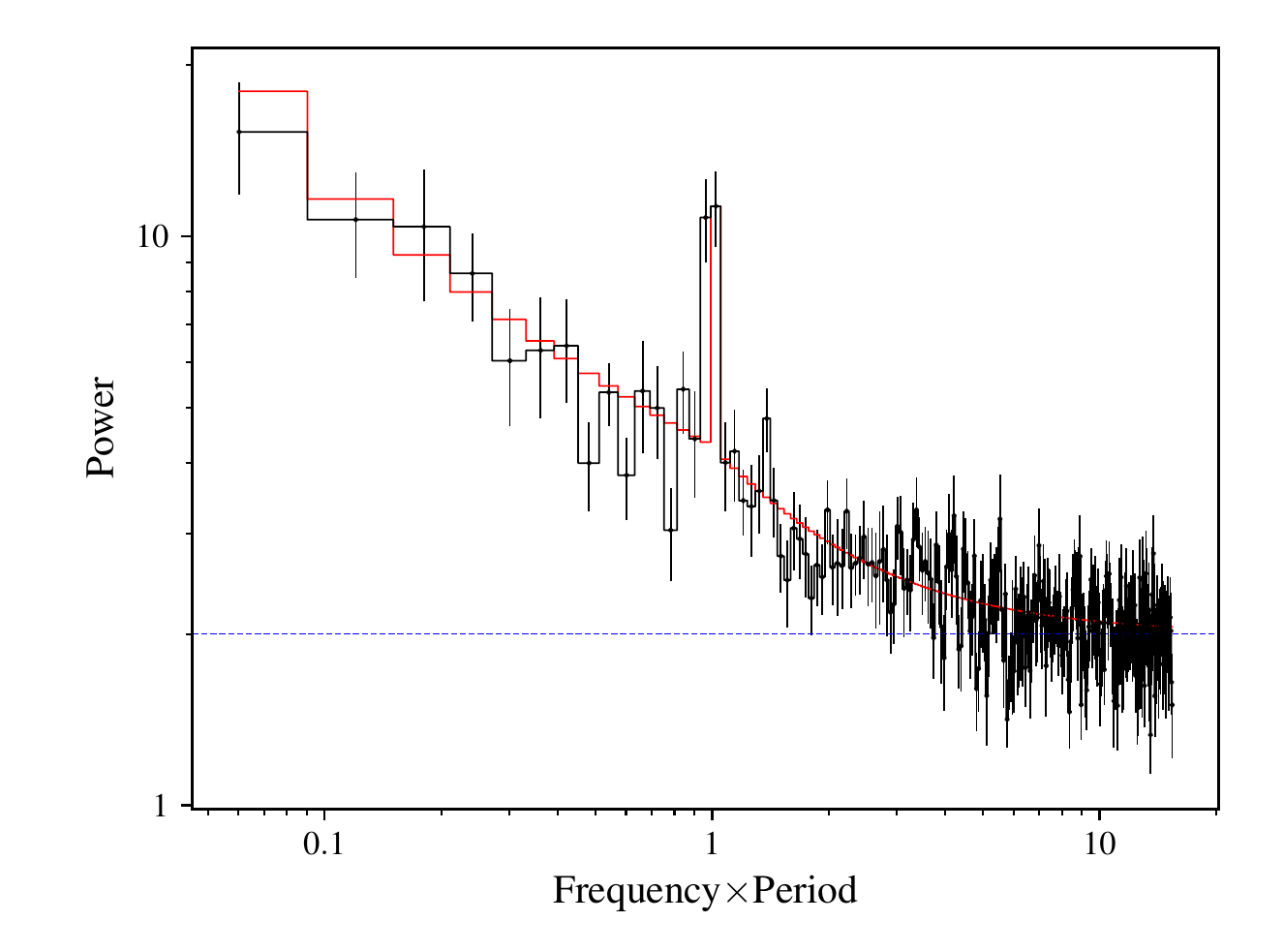}
    \caption{Leahy-normalized power spectral density of \srcnew obtained using 3-80\,keV lightcurves observed by \textit{NuSTAR} (observations 80102103002 and 90601322001). The frequency is normalized to spin frequency. Note the presence of strong red noise at lower frequencies modeled as a broken power law (red line) typical for accreting systems. The blue dashed line corresponds to expected white noise level.}
    \label{fig:psd}
\end{figure}

Besides the long spin period, also observed aperiodic variability properties can be used to probe the origin of X-ray emission. It has recently demonstrated by \cite{2020A&A...643A.173D} that unlike accreting objects, magnetars and other rotation-powered pulsars do not exhibit aperiodic variability. Therefore, we conducted analysis described in \cite{2020A&A...643A.173D} also for \srcnew. Here the data from two \textit{NuSTAR} observations with the best counting statistics (80102103002 and 90601322001) were used. As illustrated in Fig.~\ref{fig:psd}, power spectrum of \srcnew indeed shows strong aperiodic variability with broken power law type power spectrum typical of accreting objects. Indeed, when modeled as a broken power law, the relative noise amplitude (i.e. the power at the break divided by amplitude of the pulsations) estimated as described in \cite{2020A&A...643A.173D} is $35^{+13}_{-9}$\% , i.e. comparable to other accretors and much larger than in magnetars ($\le35$\%). Finally, we also inspected the light curves of the source for possible short flares, and found none.
We conclude, therefore, that together with the long pulse period and presence of a massive optical counterpart this strongly suggests that \srcnew is indeed not a magnetar but rather an accreting pulsar.

\subsection{Spectral analysis}
Besides the already mentioned \textit{Chandra} observation of the source in timed exposure imaging mode, four other observations of the source in continuous clocking (CC) mode are available. As summarized in Table~\ref{tab:obs}, three of those were done quasi-simultaneously with \textit{NuSTAR}, which enables broadband spectral analysis. Spectral products for \textit{NuSTAR} were extracted using the same procedures as described for light curves above.

As prescribed in the instruments documentation for imaging and CC modes, to extract source and background spectra from \textit{Chandra} data we used the \texttt{specextract} script, and circular or box-shaped regions centered on source and off-source with size of $2.5{\arcsec}$ and $25{\arcsec}$ respectively. 
For observations performed in CC mode, the updated source position was used to reprocess raw event files to ensure that the charge transfer inefficiency (CTI) correction is properly applied. For the observation in the imaging mode (obsid 22454) the brightness of the source was sufficient to induce significant pile-up (up to $\sim21$\% as estimated with \texttt{pileup\_map} tool). This was accounted for by including an additional pile-up model component (\texttt{jdpileup}) included in \texttt{sherpa} spectral fitting package which was used to model the source spectrum for this observation (part of \texttt{CIAO 4.12}).

Four out of five \textit{NuSTAR} observations were conducted simultaneously or quasi-simultaneously with \textit{Chandra}. In all cases \textit{Chandra} operated in continuous clocking mode, so pile-up was not an issue and was not accounted by the model. For these observations we, therefore, grouped all spectra to contain at least 25 spectral counts per energy bin and used the \texttt{Xspec v12.11.1} and standard $\chi^2$ statistics with default weighting to model the broadband spectr of the source. Initially we modeled all observations independently, however, we found that the source flux and spectrum during the the first and the last observations (obsids. 80102103002 and 90601322001 respectively) were very similar, so  these observations were modeled simultaneously for the final fit to improve counting statistics. As a result, for the broadband spectral analysis three observation groups were defined  as summarized in Table~\ref{tab:obs}.

\begin{figure}
    \centering
    \includegraphics[width=\columnwidth]{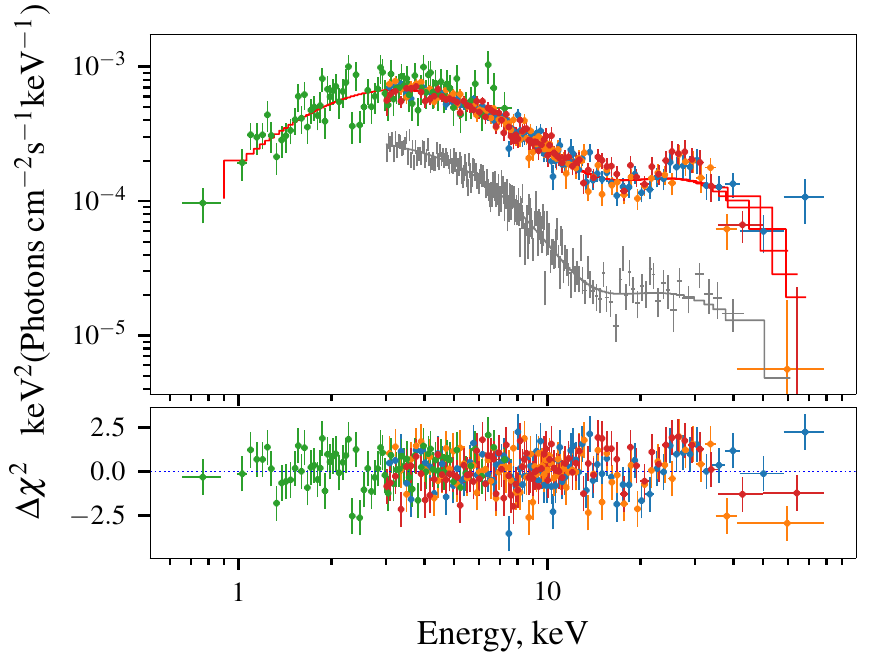}
    \caption{Unfolded spectrum of \srcnew at high flux level (spectral group A) as observed by \textit{Chandra} (green) and \textit{NuSTAR} (other colors) together with the residuals to the best fit model. For comparison spectrum of GX~304$-$1 as reported in \cite{2019MNRAS.483L.144T} is also shown (in gray; arbitrarily scaled).}
    \label{fig:spe}
\end{figure}

The broadband continuum of the source can be described by several models used to describe spectra of other low-luminosity pulsars. In particular, as shown in Fig.~\ref{fig:spe}, a two-component comptonization model \citep{2012A&A...540L...1D,2019MNRAS.483L.144T,2019MNRAS.487L..30T} provides a good approximation. More specifically, we used a model consisting of two \texttt{CompTT} components with linked seed photon temperatures and independent electron temperatures, normalizations and optical depths, for spectra of all groups A, B, and C.
In addition, we included the \texttt{tbabs} component \citep{2000ApJ...542..914W} to account for interstellar absorption. Cross-normalization constants were included to account for minor differences of the absolute flux and energy calibration, and to account for intrinsic flux variability between not strictly simultaneous individual observations. Spectra are in general well described by the model. 
The fit results are presented in Table~\ref{tab:speres} and Fig~\ref{fig:spe}. 

We note that counting statistics at higher energies is not sufficient to robustly detect or exclude the presence of possible additional features such as a cyclotron resonance scattering line.
The minor residuals around 30\,keV are likely associated with imperfect background subtraction which contains several instrumental lines around this energy and already dominates the source flux at this energy. We conclude, therefore, that the continuum is well described with a two-component Comptonization model within statistical limitations of the data. 

\begin{figure}[t!]
    \centering
    \includegraphics[width=\columnwidth]{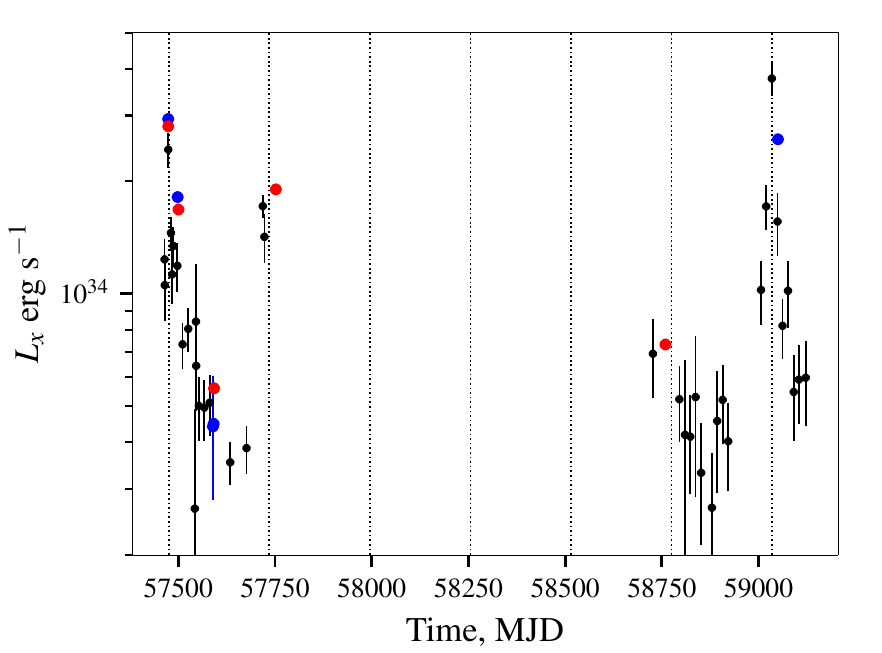}
    \caption{Bolometric long-term lightcurve of the source as observed by \textit{Swift}/XRT (black), \textit{NuSTAR} (blue) and \textit{Chandra} (red). Vertical lines indicate expected outburst timing assuming 260\,d orbital period.}
    \label{fig:longlc}
\end{figure}

To study the long-term variability of the source, we estimated the bolometric flux in each of the 45 \textit{Swift}/XRT pointed observations between MJD~57463.94 and MJD~59122.03, and in the \textit{Chandra} observations. We extracted the spectra and fitted them using the same model and parameters used for the broadband \textit{NuSTAR} and \textit{Chandra} spectra (see next paragraph) with normalization as the only free parameter. In practice, a \texttt{cflux} component was added to the best-fit model for the brightest \textit{Chandra}/\textit{NuSTAR} joint observation. The spectra were grouped to have at least one count per energy bin and modeled using the  W-statistics \citep{1979ApJ...230..274W} in the energy range of 0.4-10\,keV and 0.9-10\,keV for \texttt{PC} and \texttt{WT} modes of XRT respectively. The unabsorbed model flux was then calculated in the 0.5-100\,keV energy range to estimate bolometric luminosity of the source. The resulting lighcurve of the source is presented in Fig.~\ref{fig:longlc}.

The remaining two \textit{Chandra} observations (since there is no contemporary broadband data available) were analyzed separately similarly to \textit{Swift}/XRT data to estimate the bolometric flux of the source. To do that, we assumed the best-fit model obtained for the group A for broadband spectral analysis as described below and only considered normalization as a free parameter. We emphasize, however, that the observed spectral shape is, in fact, similar for all observations despite slight differences in the best-fit parameters, so difference in flux estimated using the best-fit models for groups A/B/C is less than 10\%, which is lower than the statistical uncertainties of individual measurements and only affects absolute flux estimate but not the shape of the light curve.

\begin{table}[]
    \centering
    \begin{tabular}{lccc}
         Parameter & A & B & C  \\
         \hline
         $N_H, 10^{22}$\,cm$^{-2}$ & $\le0.15$ & 0.13(9) & $\le0.3$ \\ 
         $T_0$, keV & 0.75(5)& 0.64(6)& 0.6(1)\\
         $kT_1$, keV & 2.1(1)& 2.1(1)& 1.6(2)\\
         $\tau_1$ & 17(2) & 19(2) & 22(4)\\
         $A_1/10^{-4}$ & 7.8(6)  & 5.1(3) & 2.0(2)\\
         $kT_2$, keV & 8.4(3) & 9.4(6) & 5.6(9)\\
         $\tau_2$ & $\ge$10 &$\ge10$ & $\ge10$\\
         $A_2/10^{-5}$ &2.7$_{-0.1}^{+2.9}$& 1.5$_{-0.8}^{+0.5}$& 0.4$_{-0.1}^{+2.0}$\\
        %  $C$ 1.0/1.06(2)/& & &\\
         $F_{-12}$ &14.4(2) & 9.6(2) & 1.5(1)\\
        %  \hline
         $\chi^2$/dof & 429.6/419 & 331.8/344 & 268.8/250\\
         
    \end{tabular}
    \caption{Best-fit results for broadband spectra of \srcnew for groups A, B, C, and for the two-component comptonization model (with components denoted by indices 1, 2) described in the text. Observed flux (i.e. not corrected for absorbtion) is given in the 3-80\,keV band in units of $10^{-12}$\,erg\,cm$^{-2}$\,s$^{-1}$. Uncertainties are reported at $1\sigma$ confidence level.}
    \label{tab:speres}
\end{table}

\section{Discussion}
 
 \begin{figure}[t!]
    \centering
    \includegraphics{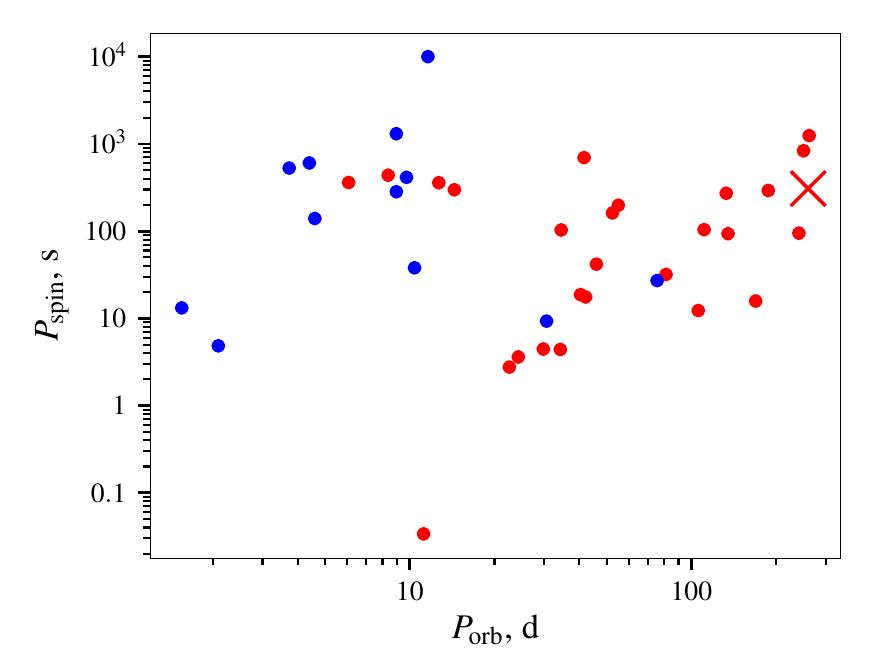}
    \caption{Location of \srcnew in the ``orbital period''--``spin period'' diagram (indicated by cross) assuming 260\,d orbital period. The red and blue points correspond to transient (BeXRB) and persistent systems respectively as reported in \cite{2006A&A...455.1165L}.}
    \label{fig:corbet}
\end{figure}

Based on the analysis of \textit{Chandra} fluxed image, the position of the initially identified soft X-ray counterpart to magnetar candidate \src has been estimated to be RA=118.9271642 Dec=$-$29.5648607 with an uncertainty of $\sim2\arcsec$. This position appears to coincide with a bright infrared and optical source CD-29~5159 known as an early type emission-line star \citep{1971PW&SO...1a...1S}. We argue, therefore, that this object is likely not the counterpart to the burst detected by BAT.
The object is detected by Gaia (DR2 source 5597252305589385984) with significant parallax of $\sim0.29(1)$\,mas. This allows to estimate the distance to the source at 3.5(2)\,kpc \citep{2018AJ....156...58B}. 
Based on the distance estimate and the observed X-ray flux, the source luminosity can be estimated at $L_{x}\sim10^{34}$\,erg\,s$^{-1}$, significantly above expectations for stellar X-ray emission, and consistent with luminosities observed from low-luminous X-ray binaries.

Pulsations at $\sim{308}$ s are significantly detected with \textit{NuSTAR}. The pulse profile is characterized by single broad peak similar to that of low-luminous accreting X-ray pulsars, i.e., X~Persei.
In addition, the power spectral density of \srcnew shows the presence at low frequencies of a strong red noise component on top the expected white noise. As recently shown by \cite{2020A&A...643A.173D} aperiodic variability is a clear indicator of accretion in compact sources.
We suggest, therefore, that \srcnew is a new low-luminous X-ray pulsar.
This hypothesis is in line with the observed broad-band X-ray spectrum of the source obtained with \textit{NuSTAR} and \textit{Chandra} consistent with a two-component Comptonization spectrum typical for low luminosity accreting pulsars (see \citealt{2012A&A...540L...1D,2019MNRAS.483L.144T,2019MNRAS.487L..30T,2020A&A...643A.173D}).

As already mentioned, the long-term variability of the source revealed by \textit{Swift}/XRT monitoring hints at a possible orbital period of $\sim260$\,d as shown in Fig.~\ref{fig:longlc}. 
Although the sampling is sparse, the duration of each of the two periods of enhanced activity around MJD~57500 and MJD~59000, each of which lasts more than 50 days, hints at a rather long orbital period. The third re-activation of the source around MJD~57750 is also consistent with this conclusion. Of course, further monitoring of the source is required to confirm it. Nevertheless, we note that the tentative orbital period identified based on recurrent flux enhancements appears to be in line with purely phenomenological expectations.
As shown in Fig.~\ref{fig:corbet}, the spin period (308\,s) and the orbital period (260\,d) allocate the source the Be X-ray binaries region of the so-called ``Corbet" diagram \citep{1986MNRAS.220.1047C}. The observed flux enhancements around the periastron are thus in line with this classification. We would like to emphasize, however, that this would be also the case for significantly shorter periods due to the large scatter of objects in the diagram so independent confirmation of the tentative period either through continued monitoring of X-ray flux or X-ray timing is essential.

We emphasize that until now the low-luminosity accretion has been mostly investigated in transient sources during their quiescent state. Peculiarly, \srcnew appears to have a rather low luminosity also during outbursts which makes it only the second source of such a kind known besides X~Persei. We note that the majority of HMXBs are actually expected to have low luminosities \citep{2014A&A...567A...7D}, and the discovery of \srcnew may mark the start of uncovering of this population hidden from instruments like \textit{Swift}/BAT. More objects like that can be expected to appear in the \textit{SRG/eRosita} survey where we expect to detect several tens of low luminous HXMBs \citep{2014A&A...567A...7D}.

\section{Conclusions}
Following the detection of a burst by \textit{Swift}/BAT, in Mar 2016, \src was classified as a new peculiar magnetar candidate. 
The discovery triggered a series of observations of the source with \textit{NuSTAR}, \textit{Chandra}, and \textit{Swift}/XRT aimed at localizing the object, searching for spin period of the neutron star and characterizing the broadband spectrum of the source. Here we have reported on the analysis of these observations. 

Surprisingly, the precise localization of the source with \textit{Chandra} suggests that the initially suggested soft X-ray counterpart to the burst observed by \textit{Swift}/BAT is unambiguously associated with the massive emission line star CD-29~5159. We argue that this, together with the long 308\,s spin period and the observed aperiodic variability of X-ray flux revealed by \textit{NuSTAR} data, rules out magnetar origin for this source. Instead we suggest the source is likely a new long-periodic low luminous HMXB which we refer to as \srcnew. Inspection of the archival \textit{Swift}/XRT data allowed to tentatively identify orbital period of $\sim260$\,d which, together with the observed spin period and classification of the optical counterpart as an emission line star suggests the BeXRB origin of the system similar to X~Persei.

On the other hand, the origin and localization of the true soft X-ray counterpart to the burst detected by \textit{Swift}/BAT (if any) remain uncertain. Only the aforementioned BeXRB was detected in the initial follow-up with \textit{Swift}/XRT, however, the later deeper \textit{Chandra} observation reveals several potential counterparts to the burst. At least two objects have no obvious counterparts in other bands and thus may be considered as an alternative magnetar candidates. Unfortunately, both objects are very faint, which precludes confirming whether it is a neutron star through the detection of pulsations or spectral analysis with the available data. Finally, one can not also fully exclude that the burst did actually originate from \srcnew. In particular, similar event might have been observed \citep{2008ATel.1731....1R,2012ApJ...744..106T} from a well-known $\gamma$-ray binary LS~I+61~303, which is also believed to be a BeXRB \citep{2006smqw.confE..52D}. We note, however, that origin of the burst is also not fully clear in this case. There are also no known $\gamma$-ray sources in immediate vicinity of \srcnew (the closest one, 4FGL~J0752.0$-$2931 is $\sim50^\prime$ away, with the positional uncertainty of only $\sim6^\prime$ at 95\% confidence level; \citealt{2020ApJS..247...33A}), so \srcnew is likely not a $\gamma-$ray binary and its connection with LS~I+61~303 is not obvious.
{We conclude, therefore, that deeper X-ray and optical observations are therefore, required, to clarify the origin of the burst and confirm it as a possible new soft gamma ray repeater.
We emphasize that also the suggested alternative counterpart might in the end be unrelated to the burst, and as such, association of the burst with a magnetar flare and classification as a magnetar candidate \src shall be considered unreliable until deeper X-ray observations are available.

\begin{acknowledgements}
This work was supported by the Russian Science Foundation (grant 19-12-00423). We thank German Academic Exchange
Service (DAAD, project 57405000) and the Academy of Finland (projects 324550, 331951) for travel grants.
JL thanks the National Natural Science Foundation of China No. 11733009, U1938103 and U2038101

\end{acknowledgements}

\vspace{-0.3cm}
\bibliography{biblio}   
\vspace{-0.3cm}

\end{document}